\documentclass[preprint, aps,prd,showpacs,nobibnotes,nofootinbib]{revtex4}
\usepackage{graphicx}
\usepackage{color, amsmath,amssymb,amsfonts, txfonts,stmaryrd}
\usepackage{setspace}
\usepackage{epsfig}
\include{psfig}
\input{epsf}

\begin{document}
\title{Domain-wall branes in Lifshitz theories}
\author{Jayne E. Thompson}
\email{j.thompson@pgrad.unimelb.edu.au} \affiliation{School of
Physics, The University of Melbourne, Victoria 3010, Australia}
\author{Raymond R. Volkas}
\email{raymondv@unimelb.edu.au} \affiliation{School of Physics, The
University of Melbourne, Victoria 3010, Australia}
\date{\today}
\begin{abstract}
We analyze whether or not Lifshitz field theories in $4+1$
dimensions may provide ultraviolet-complete domain-wall brane
models.  We first show that Lifshitz scalar field theory can admit
topologically stable domain wall solutions.  A Lifshitz fermion
field is then added to the toy model, and we demonstrate that
3+1-dimensional Kaluza-Klein zero mode solutions do {\it not} exist
when the four spatial dimensions are treated isotropically.  To
recover $3+1$-dimensional chiral fermions dynamically localized to
the domain wall, we must postulate the breaking of full 4-dimensional rotational symmetry down to the subgroup of rotations
which mix the usual 3-dimensional spatial directions and fix the extra-dimensional axis in addition to the
anisotropy between space and time.
\end{abstract}
 \maketitle
\newpage
\section{Introduction}

Ho\v{r}ava's recent attempt to solve the quantum gravity problem
\cite{Horava:2009uw, Horava:2010} by introducing higher spatial
derivative curvature terms, which break Lorentz invariance but
ameliorate graviton-loop renormalization effects, has motivated
research into other Lorentz violating theories. In particular, there
is growing interest in whether the good ultraviolat (UV) behavior of
Ho\v{r}ava-Lifshitz gravity may be transferable to certain
nonrenormalizable Yang-Mills gauge theories \cite{Iengo:2009ix,
Iengo:2010xg}. Models which include higher spatial derivative
extensions \`{a} la Ho\v{r}ava are collectively referred to as
Lifshitz field theories.

The key idea of Lifshitz field theories, in this context, is that by
explicitly breaking Lorentz invariance it is possible to write down
an action with higher-order spatial derivatives while maintaining
quadratic time derivatives and thus the unitarity of the theory. The
UV properties of the radiative corrections in the theory are
softened by the higher inverse powers of momentum appearing in the
propagator. One interesting application is to field-theoretic models
featuring extra dimensions of space. It is well-understood that the
UV behavior degrades as the number of dimensions is increased. Even
increasing the spatial dimensions from three to four causes
Yang-Mills theory to be nonrenormalizable. Thus, Lorentz-invariant
field-theory models involving extra dimensions are inherently UV
incomplete, and must be defined with a UV cutoff. But
extra-dimensional Lifshitz theories can be UV complete.

The purpose of this article is to take the standard building blocks
for a 4+1-dimensional Lorentz invariant brane world model, collated
in \cite{Davies:2007xr,Thompson:2009uk,Rubakov:1983bb}, and see if
they can be extended to construct a Lifshitz power counting
renormalizable brane world model with quartic leading order spatial
derivatives. We are principally interested in whether there is a
Lifshitz analogue to the domain-wall brane and if we can dynamically
localize fermions. These two elements (together with the
Dvali-Shifman mechanism for dynamical gauge boson localization
\cite{Dvali:1996xe}) form the backbone for a dynamically localized
domain-wall brane standard model.  Being an entree into the subject,
we restrict our analysis here to a toy model featuring just a scalar
and a fermion field.  A full theory would of course require many
more ingredients including gravity and gauge fields, but that is too
ambitious a construction to attempt in one step.

We stress that a Lifshitz domain-wall brane model would be markedly
different to previous work on extra-dimensional field theories and
valuable because preceding work in this field has been entirely on
effective field theories and by construction is only predictive up
to a UV cutoff scale \cite{Davies:2007xr,Thompson:2009uk}.

To this end we take the most general power-counting-renormalizable
action for a real Lifshitz scalar field $\phi$ living in
4+1-dimensions which is consistent with a discrete reflection
symmetry $\phi \rightarrow - \phi$, where the $Z_2$ symmetry is
necessary for generating topological boundary conditions. For this
model we demonstrate that a kink or domain-wall type of topological
defect in the Lifshitz scalar field can condense to form a brane.
However when we extend the action to incorporate fermions we find
that the standard interpretation of the 3+1-dimensional left-chiral
fermion, as the dynamically-localized zero mode in a Kaluza-Klein
tower, fails.

In isotropic $4+1$-dimensional models, Rubakov and Shaposhnikov
\cite{Rubakov:1983bb} argue that at low energies $4+1$-dimensional
spinors behave like $3+1$-dimensional left-chiral fermions localized
on the brane. Their justification is: given a solution $\Psi(x^M)$
to the isotropic $4+1$-dimensional Dirac equation, there exists a
basis for the space of continuous bounded functions of a single
coordinate, $x_5 \equiv y$, which can be used to project the field
$\Psi(x^M)$ onto a `Kaluza-Klein' tower. This Kaluza-Klein tower
contains a normalizable massless $3+1$-dimensional left-chiral
zero-mode fermion plus massive $3+1$-dimensional fermions. At low
energies only the dynamically-localized massless chiral zero mode in
this tower is kinematically allowed and the effective low-energy
dynamics carry the right phenomenological signature needed to model
the quarks and leptons of the standard model.

We take a solution to the Lifshitz $4+1$-dimensional Dirac equation
with quartic leading order spatial derivatives.\footnote{We choose
to stop at quartic order purely for simplicity.  A complete theory
including renormalizable gravity in 4+1 dimensions will require an
action containing at least order-eight derivatives.} We write this
solution as a Kaluza-Klein tower of $3+1$-dimensional spinors and we
show that this tower does not contain a massless $3+1$-dimensional
left-chiral zero mode fermion. We find that spatially isotropic
Lifshitz Dirac equations do not have zero mode solutions as part of
their Kaluza-Klein towers. To overcome this difficulty we consider
other models which do not treat all four spatial dimensions
symmetrically. We present the zero mode solution for a model where
$4+1$-dimensional Lorentz invariance is broken explicitly to ${\rm
SO}(3)$ spatial rotational invariance.

In Secs.~\ref{sec2} and \ref{sec3} we define the notation, and
Lifshitz weighted scaling dimensions convention, which we shall be
using throughout the rest of this paper. In Sec.~\ref{sec4} of this
paper we shall explicitly demonstrate that, for the minimal case of
a 4+1-dimensional model with quartic spatial derivatives, the
Euler-Lagrange equations for a general power counting renormalizable
Lifshitz scalar field action have a domain-wall brane solution. In
Sec.~\ref{sec5} we review dimensional reduction of a fermion action
via Kaluza-Klein decomposition and dynamical localization of a
candidate 3+1-dimensional fermion. We then explain precisely where
this technique fails for Lifshitz extensions. In Sec.~\ref{sec6} we
discuss alternative models featuring compact extra-dimensions and
smaller unbroken spatial symmetries. Our final section is a
conclusion.

\section{Notation}\label{sec2}

We now summarize our notational conventions for the convenience of
the reader. The lower case Latin letters will run over spatial
coordinates so that $x^i \equiv (x^1, x^2, x^3, x^5)$ and can be
lowered by contracting with the spatial metric $g_{ij}$.  Upper case
Latin letters will index  4+1-dimensional space-time coordinates so
that $x^M \equiv (t, x^i)$. In particular we distinguish the
extra-dimensional coordinate with the syntax $x^5 \equiv y$ and use
vector notation for 3-dimensional coordinate vectors,
$\vec{x}=(x^1,x^2,x^3)$.
Following standard conventions, the lower case Greek letters ${\mu}$
and ${\nu}$ are reserved for 3+1-dimensional 4-vectors so that
$x^{\mu}\equiv(t,x^1, x^2, x^3)$.  We will apply standard
terminology directly to our model without qualifying that we are
talking about a 4+1-dimensional, quartic leading order spatial
derivatives extension to the standard theory before each statement.
We will use these descriptors only when our intention is not clear
from the context. The function $\phi(x^M)$ labels the real scalar
field which condenses to form the domain-wall brane. The notation
$\Psi(x^M)$ refers to the $4+1$-dimensional fermion which we will
mode expand in a tower of fields $\psi_{n}(x^{\mu})$ corresponding
to $3+1$-dimensional massive Dirac particles. We will use the
standard symbol $\Delta \equiv g_{ij} \partial^i \partial^j $. And
by $C_b(y)$ we mean the space of continuous bounded functions with
respect to the $y$-coordinate space.

\section{Weighted Scaling Dimensions}\label{sec3}

In a $d+1$-dimensional Lifshitz type theory the space-time manifold
is foliated into a product  $R \times R^d$ with coordinates
\begin{equation}
 (t, x^i),
\end{equation}
\noindent where $i = 1,\dots, d$. The action is invariant under
$d$-dimensional spatial rotations and translations but not under
Lorentz boosts. It is useful to introduce the concept of a critical
exponent $z$ which characterizes the degree of anisotropy in the
space-time manifold. The critical exponent automatically sets the
highest power of $\Delta^z = \left(\partial_i\partial^i\right)^z$
appearing in the action for a scalar field, $\phi$, so that if we
discard all relevant operators the action possesses a rescaling
symmetry,
\begin{align}\label{eq:anisotropicscalingirreleventoperators}
  t \longrightarrow \lambda^z t, && x^i \longrightarrow \lambda x^i,  && \phi \rightarrow \lambda^{\frac{z-d}{2}} \phi.
\end{align}
Because this structure automatically adapts the loop propagators,
the correct dimensions to use when evaluating whether the theory is
power counting renormalizable are the weighted scaling dimensions in
which
\begin{align}
\label{eq:weightedscalingdimensions} \left[t\right]_s = - z, &&
\left[ x^i\right]_s= -1, &&\left[\phi\right]_s = \frac{d-z}{2}.
\end{align}
\noindent Hereinafter by $[\tau]_s$ we mean the weighted scaling
dimensions of $\tau$ which are not to be confused with the usual
mass dimensions. Lorentz invariance corresponds to $z = 1$.

To make expressions more compact we shall write $d+1_z$-dimensions
for a foliated space-time manifold with weighted scaling
characterized by (\ref{eq:weightedscalingdimensions}) and in the
isotropic space-time case we will drop the `1' subscript.

Ongoing research is critically examining whether coupling constant
running can plausibly cause Lorentz symmetry to emerge as an
accidental symmetry in the infrared limit of  $z \neq 1$ Lifshitz
models \cite{Iengo:2009ix,
Iengo:2010xg,Anselmi:2008bq,Anselmi:2008bs,Anselmi:2008bt,Chen:2009ka}.
The recent insights in Ref.~\cite{Horava:2010} now make it more
plausible that the infrared limit of a suitably-defined version of
Ho\v{r}ava-Lifshitz gravity can closely resemble pure general
relativity, so we can hope that some similar progress will be made
for emergent relativity in general.  For the purposes of this paper
we shall adopt the optimistic stance, though we recognize that the
emergence of Lorentz invariance in the infrared without fine-tuning
is not a given.

\section{domain-wall Brane}\label{sec4}

\subsection{The existence of a domain-wall Brane}\label{sec4}

We start by defining a domain-wall brane.  It is a kink-like
topological defect which defines a (finite thickness)
3+1-dimensional hyperplane to be identified with our universe. The
topological defect is provided by a real scalar field with an action
$S_{\phi}$ such that the lowest energy density solution to the
classical equations of motion, which asymptotically approaches
distinct vacua, related by a discrete reflection symmetry,  at
opposite extremities of the extra-dimension, is a solitary wave.

In this section we isolate the real scalar field terms in the
action.\footnote{It is physically justifiable to solve the scalar
field dynamical equations, and then work out the motion of fermions
propagating in the scalar field background because in the framework
of domain-wall brane models the fermions are incorporated as a
perturbative mode expansion about an empty vacuum state.}  We work
under the assumption that there is an equivalence relation on the
space of all actions such that actions $S_1$ and $S_2$ are
identified if they contain the same terms up to global boundary
terms. By consistently following this principle we will discard all
terms in the equations of motion which arise from boundary effects.
Therefore from an empirical perspective actions belonging to the
same equivalence class will be phenomenologically indistinguishable.

To create a domain-wall brane our action must exhibit a discrete
reflection symmetry $\phi (x^M) \rightarrow - \phi (x^M)$. To
simplify the model we choose to work in flat
space-time.\footnote{Incorporating gravitation will be far from
trivial, but not hopeless. First, one would have to face the
difficulties \cite{difficulties} that have been identified in the
original versions of 3+1-dimensional Ho\v{r}ava gravity
\cite{Horava:2009uw}. Reference \cite{Horava:2010} provides a way
forward.  One would then need to develop a 4+1-dimensional
extension, which will require going to a $z=4$ theory, and then
establish that the brane tension causes the domain-wall to be a
preferred observation hyperplane from which gravity appears to be
3+1-dimensional, and that the effective gravity theory on the wall
is sufficiently close to general relativity.  For domain-wall branes
with 4+1-dimensional general relativity, the answer is provided by
the Randall-Sundrum \cite{Randall:1999ee, Randall:1999vf} warped
metric solution which involves pasting together two semi-infinite
regions of anti-de Sitter space-time with a matching junction
condition at the location of the brane. This configuration can be
used to dynamically localize a 3+1-dimensional Kaluza-Klein zero
mode graviton.  Something resembling this solution would presumably
also have to exist within the putative 4+1-dimensional Ho\v{r}ava
gravity theory. The most straight-forward approach would be to
verify anti-de Sitter space-time solutions in \cite{Horava:2010}
could be extended to the putative 4+1-dimensional Ho\v{r}ava gravity
theory before looking for an analogue to the Randall-Sundrum warped
metric ansatz. These challenging problems are well beyond the scope
of this paper.} We consider the most general action for $\rm{R}
\times \rm{R}^4$ with critical exponent $z = 2$ which is consistent
with the above and in the methodology of weighted scaling dimensions
is renormalizable with a full complement of relevant and marginal
operators:
\begin{eqnarray}
 S_{\phi} &=& S_{\tiny \phi \textrm{free}} + S_{\tiny \phi \textrm{int}}, \\
 &=& \int d^5 x\, \left[ \left(\frac{1}{2} (\partial_t \phi)^2 -\frac{a^2}{2\Lambda^2} (\Delta \phi)^2  -\frac{c^2}{2}(\partial_i \phi)^2 \right) + \left(- \frac{g_2}{2} \phi^2  - \frac{g_4}{4\textrm{!} \Lambda} \phi^4 - \frac{b}{4\Lambda^3} (\phi \partial_i \phi)^2  - \frac{g_6}{6\textrm{!} \Lambda^4} \phi^6 \right) \right],\notag
\end{eqnarray}
\noindent where $\Lambda$ is being used to keep track of the natural
mass dimensions in isotropic space-time. We have absorbed the
coupling constant in front of $(\partial_t \phi)^2$ into a rescaling
of the field $\phi(x^M)$. In the weighted scaling dimensions of our
theory we have
\begin{align}
\left[t\right]_s  = -z = -2, && \left[x^i\right]_s = -1, &&
\left[\phi\right]_s = \frac{d-z}{2} = 1.
\end{align}
If we set $\left[\Lambda\right]_s = 0$ then the coupling constants
have weighted scaling dimensions:
\begin{align}
\left[a^2\right]_s = 2z - 4 = 0,  && \left[c^2\right]_s = 2z - 2 = 2, \notag \\
\left[b\right]_s = 3z - d -2  = 0,  &&  \left[g_n\right]_s = d + z -
\frac{n(d-z)}{2} = 6 - n.
\end{align}
From the above we confirm that $4+1_2$-dimensional Lifshitz scalar
field theory with a $\phi^6$ leading order potential is power
counting renormalizable. Note that the parameter $c$ which plays the
role of the maximum obtainable velocity in the free particle IR
dispersion relation, becomes a running parameter in the quantized
theory.
It would thus be misleading to absorb the energy-scale-dependent
$c^2$ by a relative rescaling of spatial and temporal coordinates.

The Klein-Gordon equation obtained from this action using the
principle of stationary action is
\begin{equation}\label{eq:KleinGordon}
\partial_t^2 \phi +\frac{a^2}{\Lambda^2}\Delta^2 \phi  - c^2 \Delta \phi - \frac{b}{2\Lambda^3} \phi^2 \Delta \phi - \frac{b}{2\Lambda^3} \phi\partial_i\phi\partial^i\phi + \frac{g_6}{5!\Lambda^4} \phi^5 + \frac{g_4}{3!\Lambda}\phi^3 + g_2\phi = 0.
\end{equation}
The domain-wall is the static solitary wave solution to the
Klein-Gordon equation which is isotropic in three spatial directions
and interpolates between distinct minima of the potential as a
function of $y$.

\begin{figure}
\centering
 \includegraphics[width=0.9\textwidth]{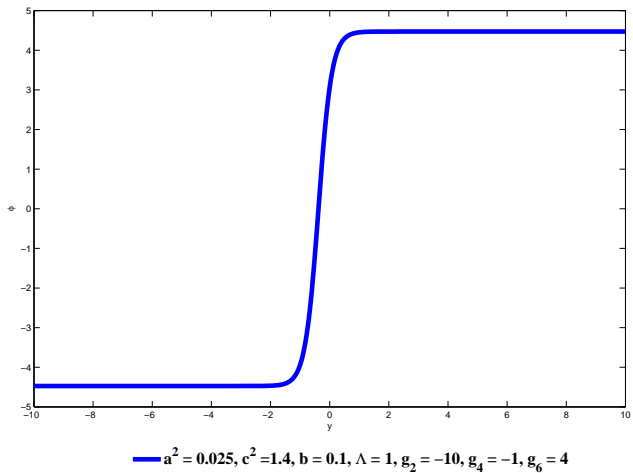}
 \caption{The graph displays a numerical domain-wall brane solution to the Lifshitz Klein-Gordon equation (\ref{eq:KleinGordon}) for a choice of parameters which do not belong to the slice defined in (\ref{eq:kinksolutionparameters}).}
\label{numericalkink}
\end{figure}

Numerical domain-wall brane solutions to the Klein-Gordon equation
have been found for a wide range of parameters. In
Fig.~\ref{numericalkink} we give an  example of a numerical
domain-wall brane configuration. In addition we can give an explicit
example of an analytic domain-wall brane,
\begin{equation}\label{eq:kinksolution}
\phi = v\textrm{tanh} (uy),
\end{equation}
\noindent which is a solution to the Klein-Gordon equation for:
\begin{equation}\label{eq:kinksolutionparameters}\begin{gathered}
v = \sqrt{-\frac{5g_4\Lambda^3}{g_6} + \frac{\sqrt{5}\sqrt{-6g_6 g_2 + 5 g_4^2\Lambda^6}}{g_6}}, \qquad u = \sqrt{\frac{bv^2}{32a^2\Lambda} - \frac{\sqrt{45b^2v^4\Lambda^2 - 16a^2g_6^4\Lambda^2}}{96\sqrt{5}a^2\Lambda^2}},\\
\textrm{and}\hspace{0.5 cm} g_6v^4 +10\Lambda(-6bu^2v^2 +
\Lambda(48a^2u^4 + \Lambda(v^2g_4 - 12c^2u^2\Lambda))) = 0.
\end{gathered}\end{equation}
\noindent We clarify that this is not a fine-tuning condition, it is
merely a prototypical example of the kink solutions that exist for a
large region of parameter space. The numerical solution depicted in
Fig.~\ref{numericalkink} corresponds to a point which is not on this
slice. Because we want the solutions described by
(\ref{eq:kinksolution}) and (\ref{eq:kinksolutionparameters}) to be
real valued and because $v$ must be a minimum of the potential for
(\ref{eq:kinksolution}) to be a solitary wave, we can bound the
allowed region of parameter space by the inequalities $\Lambda,
g_6,-g_2, -g_4 > 0$. Indeed none of the results presented in this
paper are contingent on providentially choosing parameters which
satisfy Eq.~(\ref{eq:kinksolutionparameters}).

This establishes the existence of a domain-wall brane for $R \times
R^4$ space-time with anisotropic scaling characterized by critical
exponent $z = 2$.

\subsection{Stability of the domain-wall Brane}\label{sec4}
The Lifshitz domain wall is topologically stable for the same reason
the usual domain wall is stable: the enforced discrete symmetry,
when spontaneously broken, produces disconnected vacua which serve
as the boundary conditions for the domain wall solution. The kink is
a mapping from the boundary of the real line, parameterizing the
extra dimension, onto the disconnected manifold $\{-v, v\}$. This
mapping is homotopically non-trivial.  Thus, the domain wall is
prevented from decaying to the lower-energy spatially-homogeneous
vacua $\phi(x^M) = \pm v$.  Intuitively, maintaining finite energy
density forces the solution to asymptotically approach minima of the
potential. Once we have fixed $\lim_{y \rightarrow \infty} \phi(y) =
v,$ and $\lim_{y \rightarrow - \infty} \phi(y) = -v$ there is no way
to continuously deform the solution while passing through
intermediary states which also have boundary values pinned either at
$+ v$ or $-v$ and arrive at a solution for which $\lim_{y
\rightarrow \pm \infty} \phi(y) = v$ or $\lim_{y \rightarrow \pm
\infty} \phi(y) = -v$.

We have not analytically shown that our kink domain-wall brane is the lowest energy solution to the Euler Lagrange equations satisfying these boundary conditions.

In the following discussion we guarantee the kink solution has finite 3+1-dimensional energy density (energy per unit of 3-dimensional volume). Furthermore we show that the kink is stable under perturbations corresponding to a contraction or dilation of the transverse wall width. This suggests the kink is stable because perturbations which alter the profile in any of the $\vec{x}$-coordinate directions are forbidden by ${\rm SO}(3)$-rotational invariance. Our background kink solution is invariant under 3-dimensional spatial rotations and there is no preferred direction in the hyperplane orthogonal to the $y$-coordinate, so it does not make sense to say a perturbation has formed along a specific direction in the $\vec{x}$-coordinate space because all directions are relative to an arbitrary choice of reference axis \cite{Pogosian:2000xv}.

We are left with conjecturing about perturbations which do not deform the kink in any direction perpendicular to the bulk coordinate axis and do not correspond to rescaling the width of the domain wall. This means that we can only have perturbations which locally deform the kink according to some nonlinear dependence on $y$. These are allowed because translational invariance is spontaneously broken by the condensation of the kink. A full perturbative stability analysis is beyond the scope of this paper.


Our analytic kink (\ref{eq:kinksolution}) has finite 3+1-dimensional energy density given by

\begin{equation}
\sigma = \frac{v^2 (-23g_6 v^4 + 120 \Lambda^2(24 a^2 u^4 + \Lambda(v^2 (3b u^2 - 5 g_4) -45 g_2 \Lambda + 30 c^2 u^2 \Lambda)))}{5400 u \Lambda^4} < \infty.
\end{equation}

\noindent
Naturally if we integrate this quantity with respect to the 3-dimensional volume element to obtain the total energy the answer will be infinite. This is the statement that domain-wall branes are technically only solitons in 1+1-dimensions.

In fact in isotropic 4+1-dimensional theories Derrick's theorem \cite{Derrick:1964} implies that all static non-homogeneous\footnote{ It should be qualified that the flat vacuum solutions $\phi = \pm v$ which have trivial $x^i$-dependence evade Derrick's theorem.} solutions to the Klein-Gordon equation have infinite total energy, independent of the precise set of self interaction terms in the action. Intuitively Derrick's theorem arises because a hypothetical static, stable, finite energy density solution, $\phi_0(x^i)$, to the Euler Lagrange equations would locally minimize the energy functional. However if we contract the soliton slightly, corresponding to $\phi_0(x^i) \rightarrow \phi_0(k x^i)$ where $k \approx 1$, then we find the energy functional scales according to
\begin{equation}
E[\phi_0(kx^i)] = \frac{2k^2 -1}{k^4} E[\phi_0(x^i)]
\end{equation}
\noindent
and will decrease as $k$ increases. This contradicts the hypothesis that $E[\phi_0(x^i)]$ is minimal.

In Lifshitz theories the higher spatial derivatives change the scaling properties of the energy functional. Effectively the higher spatial derivatives stabilize these perturbations and Derrick's theorem can be evaded. This means we cannot rule out finite total energy solution to the Klein-Gordon equation in Lifshitz theories. However a finite total energy solution cannot be homogeneous on 3+1-dimensional hyperplanes, so if we wish to avoid breaking 3-dimensional spatial rotational invariance then it will have to be radially symmetric and will therefore correspond to a point-like topological defect. This configuration cannot provide a 3+1-dimensional brane. We find the prospect fascinating for other reasons but have delayed a full investigation to future work.

The higher spatial derivatives in Lifshitz theories also modify the virial theorem. For a 1+1-dimensional, solitary wave solution $\phi_0(y)$ to the Klein-Gordon equation (\ref{eq:KleinGordon}), there is a relation between the contribution to the energy density from the gradient of the field, and the potential energy density. It can be derived from the total 3+1-dimensional energy density
\begin{eqnarray}
\sigma[\phi_0(y)] = \int dy\,[ \underbrace{\frac{a^2}{2\Lambda^2} (\partial_y^2 \phi_0)^2}_{S_1} + \underbrace{\frac{c^2}{2}(\partial_y \phi_0)^2 }_{S_2}  +  \underbrace{\frac{b}{4\Lambda^3} (\phi_0 \partial_y \phi_0)^2 }_{S_3} \notag \\  + \underbrace{\left[\frac{g_2}{2} \phi_0^2  + \frac{g_4}{4\textrm{!} \Lambda} \phi_0^4    + \frac{g_6}{6\textrm{!} \Lambda^4} \phi_0^6 \right]}_{S_4}] = S_1+ S_2 + S_3 +S_4,
\end{eqnarray}
\noindent
by rescaling the solution $\phi_0(y) \rightarrow \phi_0(ky)$ and computing $\left.\frac{\partial\, \sigma [\phi_0(ky)]}{\partial k}\right|_{k=1} = 0$. This result is a direct consequence of requiring any static solitary solution $\phi_0(y)$ to extremize the energy density. Here the virial relation is:
\begin{equation}
3 S_1 + S_2 + S_3 = S_4.
\end{equation}
Using this information in the second derivative $\left.\frac{\partial^2\, \sigma [\phi_0(ky)]}{\partial k}\right|_{k=1}$ we find that
\begin{equation}
 \left. \frac{\partial^2\, \sigma [\phi_0(ky)]}{\partial k}\right|_{k=1} = 18 S_1 + 2 S_2 + 2S_3.
\end{equation}
\noindent
Provided $b > 0$ the integrands of $S_1, S_2$ and $S_3$ are all perfect squares, and we recognize that this quantity must be strictly positive. By corollary a 1+1-dimensional, solitary wave solution $\phi_0(y)$ to the Klein-Gordon equation (\ref{eq:KleinGordon}) is stable under perturbations corresponding to a contraction or dilation of the transverse wall width.

\section{Fermions}\label{sec5}

We begin by carefully reviewing why Lorentz-invariant domain-wall
branes dynamically localize a massless chiral 3+1-dimensional
fermion, and then explain why this phenomenon is impossible in a
Lifshitz theory with complete spatial isotropy.

In Lorentz-invariant 4+1-dimensional brane-world models the
4+1-dimensional Dirac equation is
\begin{equation}
\label{eq:4+1-dimensionaldirac} \left[i\Gamma^M\partial_M  -  g
\phi(y) \right]\Psi(x^N) = 0,
\end{equation}
\noindent where $\Gamma^M =(\gamma^{\mu}, -i \gamma^5)$. A solution
to Eq.~(\ref{eq:4+1-dimensionaldirac}) also satisfies the
Klein-Gordon equation
\begin{equation}\label{eq:isotropicKleinGordon}
\left(\partial_t^2 - \vec{\nabla^2} -\partial_y^2 + g^2\phi(y)^2
+\gamma^5 g \left(\partial_y{\phi}(y)\right) \right)\Psi(x^M) = 0.
\end{equation}
There exist complete bases $\{f_{nL/R}(y)\}$ for $C_b(y)$ which can
be used to project the solution $\Psi(x^N)$ to
(\ref{eq:4+1-dimensionaldirac}) and (\ref{eq:isotropicKleinGordon})
onto towers of 3+1-dimensional chiral spinors $\{\psi_{n
L/R}(x^{\mu})\}$
\begin{eqnarray}\label{eq-fermionspectrum}
\Psi_i(x,y)
 &=& \Psi_{0 L}(x^M) + \sum\hspace{-0.5 cm}\int_{n > 0 } \{ \Psi_{nL}(x^M) + \Psi_{n R}(x^M) \},\notag\\
&=&  f_{0L}(y)\psi_{0L}(x^{\mu}) + \sum\hspace{-0.5 cm}\int_{n > 0 }
\{f_{n L}(y)\psi_{n L}(x^{\mu})+f_{n R}(y)\psi_{n R}(x^{\mu})\},
\end{eqnarray}
\noindent where we have introduced the notation $\Psi_{nL/R} (x^M) =
f_{n L/R}(y)\psi_{n L/R}(x^{\mu})$. We will also write
$\Psi_{n}(x^M) =  \Psi_{n L}(x^M) + \Psi_{n R}(x^M)$. There are as
many decompositions of this type as there are complete sets of
continuous bounded functions of $y$.  But, for two reasons, there is
one basis that is special. First, for this basis the $\Psi_n(x^M)$
appearing in (\ref{eq-fermionspectrum}) are independent solutions to
the Dirac equation (\ref{eq:4+1-dimensionaldirac}) which satisfy
orthonormality conditions in a rigged Hilbert space. Second, each
3+1-dimensional spinor in this tower satisfies the 3+1-dimensional
Dirac equation for a particle with mass $m_n$:
\begin{eqnarray}\label{eq:3+1-dimensionaldirac}
i \gamma^{\mu}\partial_{\mu}\psi_{nL} (x^{\mu}) &=& m_n\psi_{nR}(x^{\mu}), \notag \\
i \gamma^{\mu}\partial_{\mu}\psi_{nR} (x^{\mu}) &= & m_n
\psi_{nL}(x^{\mu}).
\end{eqnarray}
\noindent The 3+1-dimensional left-chiral zero-mode
$\psi_{0L}(x^{\mu})$ spinor satisfies
(\ref{eq:3+1-dimensionaldirac}) with $m_0 = 0$.  Equation
(\ref{eq-fermionspectrum}) does not contain a right-handed massless
3+1-dimensional spinor. Directly calculating the form of $f_{0R}(y)$
when $\psi_{0R}(x^{\mu})$ is a massless 3+1-dimensional right-chiral
fermion and $f_{0R}(y)\psi_{0R}(x^{\mu})$ is a solution to
(\ref{eq:4+1-dimensionaldirac}) reveals that $f_{0R}(y)$ does not
belong to $C_b(y)$ and hence can not form part of a basis for this
space.

From Eq.~(\ref{eq:3+1-dimensionaldirac}) it follows that each
$\psi_{nL}(x^{\mu}) + \psi_{nR}(x^{\mu})$ satisfies a
$3+1$-dimensional Klein-Gordon equation
\begin{eqnarray}\label{eq:3+1-dimensionalkleinGordon}
\left(\partial_t^2 - \vec{\nabla}^2 \right)\psi_{nL}(x^{\mu}) &=& m_n^2 \psi_{nL}(x^{\mu})\notag\\
\left(\partial_t^2 - \vec{\nabla}^2 \right)\psi_{nR}(x^{\mu}) &=&
m_n^2 \psi_{nR}(x^{\mu}).
\end{eqnarray}
\noindent If we use unbroken 3+1-dimensional Poincar\'{e} invariance
to expand $\psi_{nL}(x^{\mu}) + \psi_{nR}(x^{\mu})$ in terms of
plane waves then we find that the dispersion relation describing the
propagation of the 3+1-dimensional spinor $\psi_{nL}(x^{\mu}) +
\psi_{nR}(x^{\mu})$ with energy-momentum 4-vector ${p}_n =(\omega_n,
\vec{p}_n)$ is
\begin{equation}
\omega_n^2 - \vec{p}_n \cdot \vec{p}_n = m_n^2.
\end{equation}
From the perspective of a 3+1-dimensional observer these particles
can now be given the interpretation of propagating free particles
with masses $m_n$ which transform according to a spin-$1/2$
representation of an embedded 3+1-dimensional Lorentz space-time
symmetry. For each mode $\Psi_n (x^M)$ which appears in
(\ref{eq-fermionspectrum}) a 3+1-dimensional observer will see a
resonance in the detector at energy $m_n$ caused by what he
perceives to be a massive fermion $\psi_{n L}(x^{\mu}) + \psi_{n
R}(x^{\mu})$. At low energies only the left-chiral zero-mode
$\psi_{0L}(x^{\mu})$ will be detectable.

We choose to work with this special basis because we have a physical
interpretation for the individual modes $\Psi_n (x^M)$.  This
interpretation allows us to argue that at low energies there is a
candidate 3+1-dimensional massless chiral fermion and explain why
parity is broken (the kink localizes a zero mode of one chirality
only).

We find this special basis and the allowed masses, $m_n$, of the
3+1-dimensional spinors present in (\ref{eq:4+1-dimensionaldirac})
by solving the eigenvalue problem:
\begin{equation}\label{eq:eignevalueproblem}
\left[-\partial_y^2 + g^2\phi(y)^2 \pm g\partial_y{\phi}(y)\right]
f_{n L/R}(y) = m_n^2 f_{n, L/R}(y).
\end{equation}
Sturm-Liouville theory determines the existence and completeness of
eigensystems generated by (\ref{eq:eignevalueproblem}).

We can check these conditions are consistent and that the
eigenfunctions in (\ref{eq:eignevalueproblem}) are the correct
'special basis' functions $\{f_{n L/R}(y)\}$ to use in
(\ref{eq:4+1-dimensionaldirac}). To do this we substitute
$\Psi_{n}(x^M)$ into the 4+1-dimensional Dirac equation and use
(\ref{eq:3+1-dimensionaldirac}) to simplify. After isolating the
coefficients of the 3+1-dimensional spinors $\psi_{nL}(x^{\mu})$ and
$\psi_{nR}(x^{\mu})$, which correspond to independent degrees of
freedom, and setting each coefficient equal to zero independently we
arrive at:
\begin{eqnarray}
\left[- \partial_y - g\phi(y)\right]f_{nL}(y) &=& m_n f_{nR}(y),\notag\\
\left[ \partial_y - g\phi(y)\right]f_{nR}(y) &=& m_n f_{nL}(y).
\end{eqnarray}
\noindent Uncoupling this first order system automatically generates
the two second order differential equations in
(\ref{eq:eignevalueproblem}). We can also show that each mode
$\Psi_n(x^M)$ satisfies the Klein-Gordon equation
(\ref{eq:isotropicKleinGordon}) by using separation of variables and
Eqs.~(\ref{eq:eignevalueproblem}) and
(\ref{eq:3+1-dimensionalkleinGordon}).

We need to look at how the situation changes in the
$4+1_2$-dimensional case.

First we must clarify what we mean by a fermion in
$4+1_2$-dimensional space-time. We consider $\Psi(x^M)$ to be a
spinorial wave function on $R \times R^4$ which transforms under
$\rm{SO}(4)$ spatial rotations $x^j \rightarrow x'^j = O^j_k x^k
\approx (\delta^{j}_{k} -\epsilon^{j}_{ki}\theta^i) x^k$ according
to

\begin{equation}
\Psi(x^M) \longrightarrow S(O) \Psi (x^M), \hspace{1.5 cm} S(O) =
e^{\omega^{jk}\Sigma_{jk}} \end{equation}

\noindent where for convenience  $\omega^{jk} =
\epsilon^{jki}\theta_i$ and $\Sigma_{jk} =
-\frac{i}{4}[\Gamma_j,\Gamma_k]$. As we are talking about
4-dimensional spatial rotations here the $\Gamma^j \in \{\gamma^1,
\gamma^2, \gamma^3, \gamma^5\}$.

In the IR limit we are assuming the theory exhibits an accidental
Lorentz symmetry. This Lorentz group incorporates a 3+1-dimensional
subgroup given by the subset of ${\rm SO}(3)$ spatial rotations
augmented by boosts in the 3+1-dimensional coordinate space which
are a symmetry of the low-energy Lagrangian. Chirality will be
restored in this low-energy Lorentz-invariant theory provided we can
localize a Kaluza-Klein zero-mode fermion and provided this zero-mode fermion is an eigenstate of the $\gamma^5$ operator which
commutes with the 3+1-dimensional Lorentz subgroup.

We interpret the $4+1_2$-dimensional Dirac equation, for
$\Psi(x^M)$, to be an ${\rm SO}(4)$ covariant equation which will
putatively flow towards (\ref{eq:4+1-dimensionaldirac}) in the low
energy limit. Therefore the $4+1_2$-dimensional Dirac equation will
contain all the operators present in
(\ref{eq:4+1-dimensionaldirac}). To suppress perturbative fermion
loop diagram contributions to transition amplitudes we must
incorporate second order spatial derivatives. We choose to do this
in such a way that the propagation of the free $4+1_2$- dimensional
scalar boson and free $4+1_2$-dimensional fermion will be described
by the same dispersion relation. This means a solution to the free
field $4+1_2$-dimensional Dirac equation also satisfies a
Klein-Gordon equation which contains the same differential operators
as the Klein-Gordon equation for the free scalar field.  We will
absorb the coefficient in front of the temporal derivative into a
rescaling of the field $\Psi(x^M)$. Therefore the minimal structure
necessary for a viable Dirac equation is:
\begin{equation} \label{eq:anisotropic4+1-dimensionalDirac}
\left[i\Gamma^0\partial_t + \beta i\Gamma^i\partial_i    - \alpha
\Delta  - g \phi(y)\right] \Psi(x^M) = 0.
\end{equation}
The associated Klein-Gordon equation is
\begin{eqnarray}
\label{eq:egg}
0 &=&\left[\partial_t^2  - \beta^2 \Delta + \left(\alpha \Delta + g
\phi(y)\right)^2 - g \beta \gamma^5 \left(\partial_y \phi(y)\right)
\right]\Psi(x^M).
\end{eqnarray}
\noindent Comparing the above equation to (\ref{eq:KleinGordon})
will confirm that the free field kinetic operators are the same.

We will argue that for our $z= 2$ model, the Kaluza-Klein `zero
mode' $f_{0L}(y)\psi_{0L}(x^{\mu})$ is not a solution to the
$4+1_2$-dimensional Dirac equation. It is therefore not present in
the Kaluza-Klein tower expression (\ref{eq-fermionspectrum}) and can
not provide a candidate $3+1_2$-dimensional left-chiral fermion.

We follow the logic: If $f_{0L}(y)\psi_{0L}(x^{\mu})$ is a solution
to the $4+1_2$-dimensional Dirac equation then it is also a solution
of the Klein-Gordon equation (\ref{eq:egg}). Therefore by taking the
contrapositive of this statement: if $f_{0L}(y)\psi_{0L}(x^{\mu})$
is not a solution to (\ref{eq:egg}) then it is not a solution to the
$4+1_2$-dimensional Dirac equation. We show
$f_{0L}(y)\psi_{0L}(x^{\mu})$ is not a solution to (\ref{eq:egg}).

We do this because the operator in the Klein-Gordon equation
(\ref{eq:egg}) is diagonal when acting on a Kaluza-Klein tower
(\ref{eq-fermionspectrum}) and for the left-chiral zero mode this
operator can be simplified to an identity matrix acting on the
spinor space multiplied by a differential operator. For the
left-chiral zero mode Eq.~(\ref{eq:egg}) is
\begin{eqnarray}\label{eq:zeromodeKleinGorodon}
0 &=&\left[\partial_t^2  - \beta^2 \Delta + \left(\alpha \Delta + g
\phi(y)\right)^2 + g \beta \left(\partial_y \phi(y)\right)
\right]f_{0L}(y)\psi_{0L}(x^{\mu}).
\end{eqnarray}
It should immediately become clear that the kinetic operator in
(\ref{eq:zeromodeKleinGorodon}) contains terms like
$2\alpha\partial_y^2\vec{\nabla}^2$ which act on both the
$f_{0L}(y)$ component of this solution and the $\psi_{0L}(x^{\mu})$
component of the solution. This is different from the isotropic
4+1-dimensional Klein-Gordon equation
(\ref{eq:isotropicKleinGordon}). It means that using separation of
variables in Eq.~(\ref{eq:zeromodeKleinGorodon}) will no longer
work.

For any solution $f_{0L}(y)\psi_{0L}(x^{\mu})$ to
Eq.~(\ref{eq:zeromodeKleinGorodon}), we can use unbroken
$3+1_2$-dimensional Poincar\'{e} invariance to expand
$\psi_{0L}(x^{\mu})$ in terms of plane waves
\begin{equation}\label{eq:fourierspectrum}
\psi_{0L}(x^{\mu}) = \sum\hspace{-0.5 cm}\int d^4 p_0\, e^{-i
(\omega_0 t -  \vec{p_0} \cdot \vec{x})}\psi_{0L}(p_0^{\mu}),
\end{equation}
\noindent where the coefficient
\begin{equation}
\psi_{0L}(p_0^{\mu}) = \frac{1}{\left(2\pi\right)^4}\int d^4 x\, e^{i( \omega_0 t - i \vec{p_0}
\cdot \vec{x})} \psi_{0L}(x^{ \mu})
 \end{equation}
\noindent and the allowed values of $p_0^{\mu}$ are determined by
substituting this expansion into (\ref{eq:zeromodeKleinGorodon}). At
this stage, before finding the allowed values of $p_0^{\mu}$, we
must be clear that $p_0^{\mu}$ can not be a function of $y$.

First, if $p_0^{\mu}$ depends on $y$ then
Eq.~(\ref{eq:fourierspectrum}) will imply that $\psi_{0L}(x^{\mu})$
depends on $y$ and our interpretation of the Kaluza-Klein tower
(\ref{eq-fermionspectrum}) as a projection of the solution
$\Psi(x^M)$ to the $4+1_2$-dimensional Dirac equation down onto a
space of $3+1_2$-dimensional spinors $\psi_{n L/R}(x^{\mu})$ using a
complete basis for $C_b(y)$ is incorrect. This will mean there is a
logical inconsistency in our theory.

Second, there will be phenomenological problems for our domain-wall
brane model because when the brane is created by a soliton it has
finite width. For the analytic solution given in
(\ref{eq:kinksolution})  this width is characterized by $1/u$.  If
$p_0^{\mu}$ depends on $y$  then a particle which starts on the $y <
0$ side of the brane and propagates minutely in the $y$-direction to
the $y>0$ side will experience a sudden change in energy or momentum
$\vec{p}_0$. This is particularly true if $p_0^{\mu}$ depends in any
way on the derivative $\partial_y \phi(y)$ of the topological defect
which is large around $y=0$ because $\phi(y)$ is varying rapidly at
the location of the wall. Shrinking the width of the wall will make
the gradient of $\phi(y)$ larger around $y = 0$ and make any
dependence of $p_0^{\mu}$ on precise $y$-coordinate location more
pronounced. With this assumption we derive the dispersion relations:

\begin{eqnarray}\label{eq:momentumspacezeromodeKleinGorodon}
0 =\left[\omega_0^2  - \alpha^2 \left(\vec{p}_0^2\right)^2\right]
f_{0L}(y)  + \vec{p}_0^2 \underbrace{\left[2\alpha^2\partial_y^2  +
2\alpha g \phi(y) - \beta^2\right]}_{o1}f_{0L}(y)\notag\\
-\underbrace{\left[\left(\alpha\partial_y^2 + g\phi(y)\right)^2
-\beta^2\partial_y^2 + g \beta \left(\partial_y
\phi(y)\right)\right]}_{o2}f_{0L}(y).
\end{eqnarray}
\noindent For this dispersion relation to be $y$-independent
$f_{0L}(y)$ will have to be an eigenfunction of both $o1$ and $o2$.
Although we are free to choose any complete basis we like, there is
no function $f_{0L}(y)$ which is a simultaneous eigenfunction of
both the operators $o1$ and $o2$. Thus we have a contradiction.
There is no `zero mode' solution to the $4+1_2$-dimensional Dirac
equation and hence no candidate $3+1_2$-dimensional left-chiral
fermion at any energy regime.

We follow the source of this problem back to the higher spatial
derivative appearing among the kinetic terms in the Klein-Gordon
equation. These higher spatial derivative operators include terms
like $2\alpha^2\partial_y^2\vec{\nabla}^2$ which acted on both
$f_{0L}(y)$ and $\psi_{0L}(x^{\mu})$ and ultimately caused the
operator $o1$ to appear in
Eq.~(\ref{eq:momentumspacezeromodeKleinGorodon}). Since the problem
is due to the kinetic operator in the Klein-Gordon equation
(\ref{eq:egg}) it cannot be fixed by adding additional interaction
terms to (\ref{eq:anisotropic4+1-dimensionalDirac}). These higher
spatial derivative terms will be present in any $4+1_z$-dimensional
${\rm SO}(4)$ invariant model with $z \ge 2$. They are uniquely
fixed by ${\rm SO}(4)$ spatial rotational invariance, so the culprit
in this theory is spatial dimensional democracy.

\section{Breaking spatial isotropy}\label{sec6}

If we intend to project out effective $3+1_2$-dimensional
left-chiral fermions via Kaluza-Klein decomposition then we can not
treat all four spatial dimensions symmetrically. This immediately
prompts us to consider the alternative compact extra-dimensions
paradigm.

Consider a model where space-time is a direct-product of a
3+1-dimensional manifold, ${\cal M}^4$, and a compact
extra-dimension with an orbifolding symmetry  (the extra-dimension
is identified into cosets $S^1/Z_2$). Equip the space ${\cal M}^4
\times S_1/Z_2$ with a preferred foliation into constant time sheets
and impose anisotropic scaling characterized by critical exponent $z
= 2$.

As before the standard tactic for recovering 3+1-dimensional chiral
fermions is to Kaluza-Klein mode expand $\Psi(x^{\mu},y)$ using a
complete set $\{f_{nL/R}(y)\}$ of eigenfunctions of the
$y$-dependent component of the Dirac equation. The crucial
difference is that the boundary conditions are fixed by periodicity
with respect to $y$ of the wave-function and the transformation of
the spinor under the orbifolding symmetry. In the isotropic
space-time case these boundary conditions eliminate the right-handed
zero mode profile function and 3+1-dimensional chirality
discriminating physics is again an emergent low energy phenomenon.
Unfortunately the conceptual differences between compact and
infinite extra-dimensions change the boundary conditions for the
Euler-Lagrange equations rather than the form of the kinetic
operator. Because compact extra-dimension models still rely on the
existence of a separable solution $f_{0L}(y)\psi_{0L}(x^{\mu})$ to
the Klein-Gordon equation (\ref{eq:egg}) they will strike the same
problems.

The only option left is to strongly break 4+1-dimensional Lorentz
invariance to ${\rm SO}(3)$ spatial rotational invariance. If we
keep our anisotropic scaling with $z = 2$ then we can now write our
Dirac equation as
\begin{equation}\label{eq:SO(3)Dirac}
\left[i\Gamma^{0}\partial_{t} + i \beta \Gamma^i\partial_i +  \alpha
\Gamma^0\vec{\nabla}^2 + \lambda\partial_y^2 - g \phi(y)\right] \Psi
(x^M) = 0.
\end{equation}
In constructing Eq.~(\ref{eq:SO(3)Dirac}) we have deliberately
multiplied the differential operator $\vec{\nabla}^2$ by a matrix
which anti-commutes with $\Gamma^5$. This matrix can not be
$\Gamma^1,\, \Gamma^2$ or $\Gamma^3$ because we are demanding ${\rm
SO}(3)$ rotational invariance, therefore it must be $\Gamma^0$. We
need to introduce this matrix operator for two reasons. First,
because it engineers a Klein-Gordon equation which has a separable
solution of the form $\Psi_{0L}(x^M) = \psi_{0L}(x^{\mu})f_{0L}(y)$.
Explicitly the form of the Klein-Gordon equation is
\begin{equation}\label{SO(3)fermionkleingordon}
\left[-\left(i\partial_t + \alpha \vec{\nabla}^2\right)^2 -
\beta^2\vec{\nabla}^2 + \left(\lambda\partial_y^2 -
g\phi(y)\right)^2 - \beta^2
\partial_y^2-\beta\gamma^5\left(\partial_y g \phi (y)\right) \right]
\Psi (x^M) = 0.
\end{equation}
There are no operators in Eq.~(\ref{SO(3)fermionkleingordon}) which
operate on both the $x^{\mu}$ and $y$ coordinate spaces
simultaneously.

Second, to enable the effective $3+1_2$-dimensional theory to be
chiral, we need the left- and right-chiral components of the zero
mode to be independent solutions of Eq.~(\ref{eq:SO(3)Dirac}). We
can then arrange for the right-chiral profile function $f_{0R}(y)$
not to belong to $C_b(y)$ so that the right chiral `zero mode' is
excluded from the Kalza-Klein tower. We collect the
$x^{\mu}$-coordinate space differential operators in
(\ref{eq:SO(3)Dirac}) and interpret the $3+1_2$-dimensional zero
mode massless spinor to be the solution of
\begin{equation}\label{eq:masslessdiracequations}
\left[ i\gamma^{0}\partial_{t} + i \beta \left(\gamma^1\partial_{1}
+ \gamma^2\partial_{2} + \gamma^3\partial_{3} \right) +
\alpha\gamma^0\vec{\nabla}^2\right] \psi_{0L/R}(x^{\mu}) = 0.
\end{equation}
\noindent It is easy to check that $\psi_{0L}(x^{\mu})$ and
$\psi_{0R}(x^{\mu})$ will only be independent solutions to
(\ref{eq:masslessdiracequations}) when the $\vec{\nabla}^2$
differential operator term in Eq.~(\ref{eq:masslessdiracequations})
is multiplied by $\gamma^0$.

After using (\ref{eq:masslessdiracequations}) to simplify the
algebra we find that $\Psi_{0L/R}(x^M) =
f_{0L/R}(y)\psi_{0L/R}(x^{\mu})$ will be a solution to
(\ref{eq:SO(3)Dirac}) if
\begin{eqnarray}
0 & = & \lambda\partial_y^2 f_{0L}(y) - \beta \partial_y f_{0L}(y) - g\phi(y)f_{0L}(y),\notag\\
0 & = & \lambda\partial_y^2 f_{0R}(y) + \beta \partial_y f_{0R}(y) -
g\phi(y)f_{0R}(y).
\end{eqnarray}
\noindent
 If we make the substitution $f_{0L/R}(y) = e^{\pm \beta y/2\lambda} F_{0L/R}(y)$ then these equations can be converted into a time independent Schr\"{o}dinger equation for $F_{0L/R}(y)$:
\begin{equation}\label{eq:timeindependentschrodinger}
\left[-\partial_y^2 + \frac{g}{\lambda}\phi(y) +
\frac{\beta^2}{4\lambda^2} \right] F_{0L/R}(y) = 0.
\end{equation}
\noindent We observe that for the kink form of $\phi(y)$ given in
(\ref{eq:kinksolution}) and parameters belonging to the open set
$4gv\lambda > \beta^2 $, the solution $F_{0L/R}(y)$ to
(\ref{eq:timeindependentschrodinger}) describes a freely propagating
particle in the region $y < u \textrm{ arctanh}
(-\beta^2/4gv\lambda)$ which is incident on a potential barrier for
$y > u\textrm{ arctanh} (-\beta^2/4gv\lambda)$. Because all
numerical solutions have the same kink like behavior we can extend
this analysis to cover all domain-wall branes. We treat the analytic
case here because it has a closed form expression for the location
of the potential barrier in (\ref{eq:timeindependentschrodinger}).
We can use the WKB method to find an approximate solution to
(\ref{eq:timeindependentschrodinger}) and write the zero mode
profile functions in terms of our analytic solution
(\ref{eq:kinksolution}) for $\phi(y)$, as
\begin{equation}\label{eq:profiles}
f_{0L/R}(y) \approx  e^{\pm\beta y/2\lambda} \left\{
     \begin{array}{lr}
      \sqrt[4]{\frac{4\lambda^2}{ 4 \lambda g\phi(y)+ \beta^2}} e^{-\int_{u\textrm{ arctanh} (-\beta^2/4gv\lambda)}^y dy' \sqrt{  g\phi(y')/\lambda + \beta^2/4\lambda^2 }} & : y > u\textrm{ arctanh} (-\beta^2/4gv\lambda)\\
       \sqrt[4]{\frac{4\lambda^2}{-\beta^2 - 4 \lambda g\phi(y)}} e^{i\int_{u\textrm{ arctanh} (-\beta^2/4gv\lambda)}^y dy' \sqrt{-\beta^2/4\lambda^2 -  g\phi(y')/\lambda}} & : y < u\textrm{ arctanh} (-\beta^2/4gv\lambda).
     \end{array}
   \right.
\end{equation}
This approximation breaks down at the cusp of (\ref{eq:profiles}),
$y = u\textrm{ arctanh} (-\beta^2/4gv\lambda)$. Provided $\beta > 0$
the profile function $f_{0R}(y)$ grows exponentially as $y
\rightarrow - \infty$ and the right handed `zero mode' will not
belong to our Kaluza-Klein tower. In addition $f_{0L}(y)$ satisfies
$\lim_{y \to \pm \infty} f_{0L}(y) = 0 $. This follows from
substituting the asymptotic form $\lim_{y \to \pm \infty}
\phi(y)\rightarrow \pm gv$  for the analytic kink $\phi(y)$ given in
(\ref{eq:kinksolution}) into the expression for $f_{0L}(y)$
(\ref{eq:profiles}) and taking the appropriate limits. The
left-chiral profile function $f_{0L}(y)$ decays exponentially in
both directions and therefore given $\epsilon > 0$, $\exists
y(\epsilon) $ such that $| f_{0l}(y)| < \epsilon,\, \forall |y| \geq
y(\epsilon)$ and on the compact region $|y| < y(\epsilon)$ we can
argue that $f_{0L}(y)$ is continuous and hence attains a finite
maximum , thus $f_{0L}(y)$ is bounded.

We can now solve Eq.~(\ref{eq:masslessdiracequations}) to obtain the
form of the $3+1_2$-dimensional spinor $\psi_{0L}(x^{\mu})$. We
choose to write the full result for the left-chiral zero mode in
terms of the solution for $f_{0L}(y)$ given in (\ref{eq:profiles})
as:
\begin{equation}
\Psi_{0L} (x^M) = N f_{0L}(y) \left(\begin{matrix} \chi_{+}  \\
0\end{matrix}\right) e^{-i(\omega_0 t - \vec{p_0}\cdot \vec{x})},
\end{equation}
\noindent where $N$ is a normalization constant, the two component
column vector $\chi_{+}$ is the eigenstate of the helicity
operator\footnote{The helicity operator gives the component of the
spin of $\psi_{0L}(x^{\mu})$ in the direction of propagation in the
3-dimensional $\vec{x}$ coordinate space. It is defined in terms of
a vector of Pauli matrices $\vec{\sigma} = (\sigma_1,
\sigma_2,\sigma_3)$ and the particle's 3-momentum, $\vec{p}_0$, as
$\vec{\sigma}\cdot \vec{p}_0/|\vec{p}_0|$.} corresponding to
$\vec{\sigma}\cdot \vec{p}_0 \chi_+ = |\vec{p}_0| \chi_+$, the lower two components of the four component spinor are zero and the
energy and momentum of the plane wave $p_0^{\mu} = \left(\omega_0,
\vec{p}_{0}\right)$ is fixed by the dispersion relation derived by
substituting this solution into the Klein-Gordon equation
(\ref{SO(3)fermionkleingordon}) to obtain
\begin{equation}
\left(\omega_0 - \alpha \vec{p}_0^2\right)^2 - \beta^2\vec{p}_0^2 =
0.
\end{equation}
This energy-momentum relation for the $3+1_2$-dimensional subspace
no longer depends on $y$.

However if we derive the free scalar field Klein-Gordon equation
directly from a $z=2$, ${\rm SO}(3)$ invariant action for
$\phi(x^M)$ we will arrive at
\begin{equation}\label{eq:SO(3)KleinGoron}
\left(\partial_t^2 -\frac{a_1^2}{\Lambda^2}
\left(\vec{\nabla}^2\right)^2 - \frac{a_2^2}{\Lambda^2} \partial_y^4
- \frac{a_3^2}{\Lambda^2} \partial_y^2\vec{\nabla}^2  -
c_1^2\vec{\nabla}^2 -c_2^2 \partial_y^2\right) \phi(x^{M}) = 0.
\end{equation}
\noindent The domain-wall brane (\ref{eq:kinksolution}) will still
be a solution to (\ref{eq:SO(3)KleinGoron}). In fact in the
spatially anisotropic scenario where Lorentz invariance is broken
directly to $\textrm{SO}(3)$ spatial rotational invariance and the
real scalar field Klein-Gordon equation (involving self interaction
terms) has the form:

\begin{eqnarray}\label{eq:completeSO(3)KleinGoron}
\partial_t^2\phi -\frac{a_1^2}{\Lambda^2}
\left(\vec{\nabla}^2\right)^2 \phi - \frac{a_2^2}{\Lambda^2}
\partial_y^4 \phi - \frac{a_3^2}{\Lambda^2}
\partial_y^2\vec{\nabla}^2 \phi  - c_1^2\vec{\nabla}^2 \phi -c_2^2
\partial_y^2 \phi - \frac{b_1}{2\Lambda^3} \phi^2 \vec{\nabla}^2
\phi  && \notag\\  -\frac{b_1}{2\Lambda^3}
\phi\vec{\nabla}\phi\cdot\vec{\nabla}\phi - \frac{b_2}{2\Lambda^3}
\phi^2\partial_y^2 \phi - \frac{b_2}{2\Lambda^3}
\phi\partial_y\phi\partial_y\phi- \frac{g_6}{5!\Lambda^4} \phi^5 -
\frac{g_4}{3!\Lambda}\phi^3 - g_2\phi  &=& 0,
\end{eqnarray}

\noindent our solitary wave domain-wall brane
(\ref{eq:kinksolution}) will be a solution  provided the free
parameters satisfy (\ref{eq:kinksolutionparameters}) with the
identifications $a \equiv a_2,\, c \equiv c_2,\, \textrm{and } b
\equiv b_2$.

From a comparison of (\ref{eq:SO(3)KleinGoron}) and
(\ref{SO(3)fermionkleingordon}) we infer that the fermions and
bosons will now propagate according to different dispersion
relations. It is easy to check that if we insist on writing down a
Dirac equation for which the solutions $\Psi (x^M)$ are also
solutions to (\ref{eq:SO(3)KleinGoron}) for the case $a_3 = 0$ (this
will circumvent our previous problem of mixed partial derivative
operators) then we will have to use a 7-dimensional Clifford
algebra. This forces the smallest representation for our spinor
$\Psi(x^M)$ to consist of 8-component column matrices and ultimately
once this Dirac equation has been solved we find that the `zero
mode' has too many degrees of freedom to be given the interpretation
of a massless chiral fermion.

\section{Conclusion}

In this paper we have examined a $4+1_z$-dimensional Lifshitz scalar
field theories, with critical exponent $z=2$. We have demonstrated
that a topological defect in the scalar field can spontaneously
condense to form a domain-wall brane. We consider the dynamics of a
Lifshitz fermion in this background and show that a
$3+1_2$-dimensional left chiral Kaluza-Klein zero mode fermion will
become trapped on the wall only when four fold spatial rotational
invariance is strongly broken to three-fold rotational invariance.
So it is not possible to keep 4-dimensional spatial isotropy and
localize a  Kalza-Klein zero mode candidate chiral fermions in
4+1-dimensional Lifshitz domain-wall brane models. While we worked
with $z=2$ theory for simplicity, we do not expect the results to be
qualitatively different for the more realistic case of $z=4$,
necessary to have a renormalizable quantum theory of gravity in
$4+1$ dimensions.  Many challenges remain before a realistic
UV-complete domain-wall brane model could be contemplated, including
the dynamical localization of gravity in some type of
Ho\v{r}ava-Lifshitz theory, and the incorporation of
dynamically-localized gauge fields.

\subsection*{Acknowledgments}

We thank D. P. George and A. Kobakhidze for helpful discussions. In
particular we acknowledge D. P. George independently proposed that a
Lifshitz domain-wall brane model will have improved UV-behaviour
leading to a strengthening of the Dvali-Shifman conjecture for
gauge-field localization. This work was supported in part by the
Australian Research Council and in part by the Puzey Bequest to the
University of Melbourne.

\end{document}